\documentclass[12pt]{iopart}
\usepackage{epsfig}

\def\beq{\begin{equation}}
\def\eeq{\end{equation}}

\def\barr{\begin{eqnarray}}
\def\earr{\end{eqnarray}}
\def\lsim{\raise0.3ex\hbox{$\;<$\kern-0.75em\raise-1.1ex\hbox{$\sim\;$}}}
\def\gsim{\raise0.3ex\hbox{$\;>$\kern-0.75em\raise-1.1ex\hbox{$\sim\;$}}}
\def\bmat{\left( \begin{array}}
\def\emat{\end{array} \right)}

\def\dmsq{\Delta m^2}

\def\msqs{\Delta m^2_{\odot}}

\def\nue{{\nu_e}}
\def\nuebar{{\bar{\nu}_e}}

\def\nuxbar{{\bar{\nu}_x}}
\def\ebar{{\bar{e}}}
\def\xbar{{\bar{x}}}
\def\dmbar{{\overline{\dmsq_\oplus}}}

\begin{document}

\title[Earth effects on SN neutrinos at a single detector]
{Identifying Earth matter effects on supernova neutrinos
at a single detector} 

\author{Amol S.~Dighe, Mathias Th.~Keil and Georg G.~Raffelt}

\address{Max-Planck-Institut f\"ur Physik
(Werner-Heisenberg-Institut),
F\"ohringer Ring 6, 80805 M\"unchen, Germany}

\begin{abstract}
The neutrino oscillations in Earth matter introduce modulations
in the supernova neutrino spectra.
These modulations can be exploited to identify the presence of 
Earth effects on the spectra, which would enable us to put 
a limit on the value of the neutrino mixing angle $\theta_{13}$ 
and to identify whether the mass hierarchy is normal or inverted.
We demonstrate how the Earth effects can be identified 
at a single detector without prior assumptions about the 
flavor-dependent source spectra, using the 
Fourier transform of the ``inverse-energy'' spectrum of the signal.
We explore the factors affecting the efficiency of this method,
and find that the energy resolution of the detector is the most 
crucial one. 
In particular, whereas water Cherenkov detectors may need a few ten 
thousand events to identify the Earth effects, 
a few thousand may be enough at  scintillation
detectors, which generically have a much better energy resolution.
A successful identification of the Earth effects through this method
can also provide $\msqs$ to a good accuracy.
The relative strength of the detected Earth effects as a function 
of time provides a test for supernova models.

\end{abstract}


\maketitle

\section{Introduction}
\label{intro}

Our knowledge of neutrino masses and mixing parameters has been 
rapidly improving  in the past few years. 
We know that the neutrino flavors $\nu_e, \nu_\mu$ and 
$\nu_\tau$ mix among themselves, and we already have measured
the mass squared differences and two of the three mixing angles 
to a good accuracy 
\cite{Gonzalez-Garcia:2002dz}--\cite{Gonzalez-Garcia:2003qf}.
The most important unknowns
are the value of the third mixing angle, whether the neutrino mass 
hierarchy is normal or inverted, and the magnitude of the leptonic 
CP violation.
The neutrino spectra from a core collapse supernova (SN) can shed
light on the first two of these unknowns.

Neutrinos produced inside a SN core
undergo oscillations on their way out through the mantle
and envelope of the star, through the interstellar space,
and possibly even through some part of the Earth
before arriving at the detector.
The spectra of these neutrinos carry information about the 
two mass squared differences and the $\nu_e$ flavor component in the 
three mass eigenstates. 
Of course this information comes convoluted with the primary
fluxes of the neutrinos produced inside the star, and
the extraction of the oscillation parameters depends crucially
on our understanding of these primary fluxes.

The uncertainties in the calculations of the primary flux
spectra remain large, so
only some of the robust features of these spectra can be used
with confidence to extract the mixing parameters.
Even with this limitation, it has been argued 
\cite{Dighe:1999bi,Lunardini:2003eh}
that the observations of the $\nue$ and $\nuebar$ spectra at
the detectors on Earth may reveal the type of the neutrino
mass hierarchy and imply a range for the third mixing angle.
Significant modifications of neutrino spectra can take place  
if the neutrinos travel through the Earth matter before reaching
the detector, and these Earth effects can provide perhaps the most 
concrete signatures for some of the neutrino mixing scenarios
\cite{Dighe:2001rs,Lunardini:2001pb,Takahashi:2001dc}.

Any signature of neutrino oscillations depends on the 
flavor-dependent differences between the primary spectra. 
Recent studies 
\cite{Raffelt:ai,Buras:2002wt,Keil:2002in,Noon03} reveal that
these differences are much smaller than had been assumed before.
This makes the identification of Earth effects harder than 
that expected in the previous analyses, and one needs to
reevaluate the potential of the detectors from the perspective
of these ``pessimistic'' assumptions about the primary 
fluxes and spectra.

The comparison of the neutrino signals observed at two detectors 
would be the most efficient way of detecting the Earth effects.
However the detectors need to be sufficiently far apart so that the
Earth effects are different for both of them, and sufficiently large 
to observe a statistically significant signal.
Currently Super-Kamiokande is the only large detector that can measure
the neutrino energies and can detect more than a few thousands of
events from a galactic SN.

Though two detectors, both capable of measuring the neutrino
energies, is definitely the most desirable option, it is still
possible to detect Earth effects without having to measure the
energies of individual neutrinos.
Indeed, it has recently been shown that for a galactic SN, the
comparison of the number of Cherenkov photons detected at 
Super-Kamiokande and IceCube as a function of time may be
able to identify the Earth effects \cite{Dighe:2003be}. 
The location of the SN can be anywhere within a specific 70\% region of 
the sky for the effects to be observable.

The motivations to look for a way to identify the Earth effects at 
a single detector are manifold, the cancellation of
systematic uncertainties being the major one. Moreover, in
conjunction with the two-detector signature mentioned above, 
the fraction of the sky for a favorable location of the 
SN increases to about 85\% when Super-Kamiokande
is also employed as the ``single'' detector. 
For nearly 35\% of the sky fraction for the SN location, 
these two ways of detecting the Earth effects act as confirmatory 
tests of each other.
The ranges of primary neutrino flux parameters that give an
Earth effect signature through these two procedures also differs
slightly, since the two-detector method involving IceCube relies on the
measurement of integrated luminosity, whereas the one-detector 
method relies on the analysis of the spectral shape.
In this sense, the two methods are complementary to each other.

The naive method of fitting the observed neutrino signal 
for the primary fluxes and neutrino mixing parameters is 
inefficient for several reasons. The primary fluxes -- their
average energies, spectral widths as well as the magnitude of
the total fluxes -- are time dependent, and only a few rough
features can be said to be known with any confidence. One may
try to get rid of the time dependence by dividing the signal 
into several time bins, but this reduces the available statistics.
Even within each time bin, one typically has to fit for as many as 
5 parameters of the primary spectra, not counting the 
uncertainties in the neutrino mixing parameters.

However, robust identification of Earth effects can be achieved
by observing that the parameters that govern the oscillation frequency 
of the neutrinos inside the Earth are relatively well measured, and 
more importantly, completely independent of the primary neutrino
spectra. 
Therefore, the Earth effects can be identified merely by
identifying the presence of this oscillation frequency in the 
observed spectrum. 
The oscillation frequency does not change with time, thus
obviating the need for several time bins.
This oscillating component is expected to be
a small addition to the otherwise approximately blackbody
spectrum that forms the major component of the signal.
In order to extract this small oscillation component, we propose a
Fourier analysis of the spectrum, which can separate the signals
of different frequency from their superposition. We perform a
numerical simulation to explore the efficiency of this
method at a water Cherenkov detector like Super-Kamiokande,
and a large scintillation detector like the one proposed in
Ref.~\cite{lena}, which has a much better energy resolution.

Since the Earth effects can be observed only for
certain neutrino mixing scenarios, the mere identification of
these effects can already tell us whether, for example, the
neutrino mass hierarchy is normal. This is a daunting task 
even at the future long-baseline experiments. In addition,
the magnitude of the Earth effects as a function of
time gives us important information about the primary neutrino
spectra that is extremely useful for understanding the 
SN dynamics.

This paper is organized as follows.
In Sec.~\ref{osc}, we 
describe the modifications of the primary neutrino spectra
by the Earth matter effects.
In Sec.~\ref{identify}, we introduce the Fourier transform
for extracting the oscillation frequency of neutrinos
inside the Earth, and propose a procedure for identifying
the Earth effects without having to make any assumptions
about the primary neutrino spectra. 
In Sec.~\ref{resol}, we explore this method 
at scintillation and water Cherenkov detectors through
a Monte Carlo simulation.
Sec.~\ref{concl} concludes.

\section{Neutrino mixing parameters and Earth effects on the 
\boldmath{$\nuebar$} spectrum}
\label{osc}

Neutrino oscillations are now firmly established by measurements of
solar and atmo\-spheric neutrinos and the KamLAND \cite{kamland} and 
K2K \cite{k2k} long-baseline experiments.
The weak interaction
eigenstates $\nu_e$, $\nu_\mu$ and~$\nu_\tau$ are non-trivial
superpositions of three mass eigenstates $\nu_1$, $\nu_2$ and~$\nu_3$,
\beq
\nu_\alpha = U_{\alpha i} \nu_i~, \quad \quad \quad \quad   
\alpha \in \{e,\mu,\tau\}, ~i \in\{1,2,3\}~,
\eeq
where $U$ is the leptonic mixing matrix. It can be written in the 
canonical form
\beq
U = R_{23}(\theta_{23}) R_{13}(\theta_{13}) R_{12}(\theta_{12})~~,
\eeq
where $R_{ij}(\theta_{ij})$ corresponds to the rotation in the
$i$--$j$ plane through an angle $\theta_{ij}$. We have neglected
the CP violating effects, which are irrelevant for SN
neutrinos.
From a global 3-flavor analysis of all the available data, 
one finds the 3$\sigma$ ranges for the mass squared differences 
$\dmsq_{ij} \equiv m^2_i - m^2_j$ 
and mixing angles
as summarized in Table~\ref{tab:mixingparameters}.

\begin{table}[ht]
\caption{\label{tab:mixingparameters}Neutrino mixing parameters from a
global analysis of all experiments 
(3$\sigma$~ranges)~\cite{Gonzalez-Garcia:2003qf}.}
\begin{indented}
\item[]\begin{tabular}{@{}lll}
\br
Observation&Mixing angle&$\Delta m^2$ [meV$^2$]\\
\mr 
Sun, KamLAND & $\theta_{12}$ = 32$^\circ$--42$^\circ$&
$\dmsq_{21} = m^2_2-m^2_1 =$ 54--190\\
Atmosphere, K2K  & $\theta_{23} =$ 34$^\circ$--60$^\circ$&
$|\dmsq_{32}| = |m^2_3 - m^2_2|=$ 1500--3900\\
CHOOZ & $\theta_{13} {}<13^\circ$ & 
$\dmsq_{31} = m^2_3 - m^2_1 \approx \Delta m_{32}^2$\\
\br
\end{tabular}
\end{indented}
\end{table}

A SN core is essentially a neutrino blackbody source, but small
flavor-dependent differences of the fluxes and spectra remain.  
Since these differences are very small between $\bar\nu_\mu$
and $\bar\nu_\tau$, we represent both these species by $\nuxbar$.
We denote the fluxes of $\nuebar$ and $\nuxbar$ at Earth that would be
observable in the absence of oscillations by $F_\ebar^0$ and $F_\xbar^0$,
respectively. For both $F_\ebar^0$ and $F_\xbar^0$, we
assume a distribution of the form~\cite{Keil:2002in}
\begin{equation}\label{eq:spectralform}
F(E)=
\frac{\Phi_0}{E_0}\,\frac{(1+\alpha)^{1+\alpha}}{\Gamma(1+\alpha)}  
\left(\frac{E}{E_0}\right)^\alpha 
\exp\left[-(\alpha+1)\frac{E}{E_0}\right]\,,
\end{equation}
where $E_0$ is the average energy, $\alpha$ a parameter that typically
takes on values 2.5--5 depending on the flavor and the phase of
neutrino emission, and $\Phi_0$ the overall flux at the detector.
The values of the total flux $\Phi_0$
and the spectral parameters $\alpha$ and $E_0$ are different for
$\nuebar$ and $\nuxbar$. We represent these by the appropriate subscripts.
Two of the most important robust features of the primary spectra are 

(1) the energy hierarchy of the neutrino species:
$\langle E_{\nu_e} \rangle < \langle E_{\nuebar} \rangle 
\lsim \langle E_{\nuxbar} \rangle$,

(2) ``pinching'' of the spectra: $\alpha_i >2$ for all species.

\noindent 
Apart from a heavy water detector like SNO, the 
neutrino detectors are sensitive mainly to $\nuebar$ in the
SN energy range. We shall therefore concentrate only on the
$\nuebar$ spectrum in this paper.
In the presence of oscillations a $\bar\nu_e$ detector actually observes
the flux 
\begin{equation}
F_{\bar{e}}^D(E)  =  \bar{p}^D(E) F_{\bar{e}}^0(E) + 
\left[ 1-\bar{p}^D(E) \right] F_\xbar^0 (E)\,,
\label{feDbar}
\end{equation}
where $\bar{p}^D(E)$ is the $\bar\nu_e$ survival probability after
propagation through the SN mantle and perhaps part of the Earth before
reaching the detector. The bulk of the $\nuebar$ are observed 
through the inverse beta decay reaction $\nuebar p \to n e^+$.
The cross section $\sigma$ of this reaction is proportional to
$E_{\nuebar}^2$, making the spectrum of neutrinos observed at the
detector $N(E_{\nuebar}) \propto  \sigma F_{\bar{e}}^D \propto 
E_{\nuebar}^2 F_{\bar{e}}^D$.

A significant modification of the survival probability due to the
propagation through the Earth appears only for those combinations of
neutrino mixing parameters shown in Table~\ref{tab:EarthCases}.  The
Earth matter effects depend strongly on two parameters, the sign of
$\dmsq_{32}$ and the value of
$\sin^2 \theta_{13}$~\cite{Dighe:1999bi,Dighe:2001rs}.  The ``normal
hierarchy'' corresponds to $m_1<m_2<m_3$, i.e.\ $\dmsq_{32}>0$,
whereas the ``inverted hierarchy'' corresponds to $m_3<m_1<m_2$, i.e.\
$\dmsq_{32}<0$.  Note that the presence or absence of the Earth effects
discriminates between values of $\sin^2 \theta_{13}$ less or greater
than $10^{-3}$, i.e.\ $|\theta_{13}|$ less or larger than about
$1.8^\circ$.  Thus, the Earth effects are sensitive to values of
$|\theta_{13}|$ that are much smaller than the current limit.

\begin{table}[ht]
\caption{\label{tab:EarthCases}The Earth effects appear for
the indicated flavors in a SN signal.}
\begin{indented}
\item[]\begin{tabular}{@{}lll}
\br
13-Mixing&Normal Hierarchy&Inverted Hierarchy\\
\mr
$\sin^2\theta_{13} \lsim 10^{-3}$&$\nu_e$ and $\bar\nu_e$
&$\nu_e$ and $\bar\nu_e$\\
$\sin^2\theta_{13} \gsim 10^{-3}$&$\bar\nu_e$&$\nu_e$ \\
\br
\end{tabular}
\end{indented}
\end{table}

Let us consider those scenarios where the mass hierarchy and the value
of $\theta_{13}$ are such that the Earth effects appear for
$\bar\nu_e$.  In such cases the $\bar\nu_e$ survival probability
$\bar{p}^D(E)$ is given by
\begin{equation}
\bar{p}^D \approx \cos^2 \theta_{12} 
- \sin 2\bar{\theta}_{e2}^\oplus
~ \sin (2\bar{\theta}_{e2}^\oplus - 2\theta_{12}) 
~\sin^2 \left( 12.5\,\frac{\overline{\dmsq_\oplus} L }{E} \right)\,,
\label{pbar}
\end{equation}
where the energy dependence of all quantities will always be implicit.
Here $\bar{\theta}_{e2}^\oplus$ is the mixing angle between $\nuebar$
and $\bar{\nu}_2$ in Earth matter while $\overline{\dmsq_\oplus}$ is
the mass squared difference between the two anti-neutrino mass
eigenstates $\bar{\nu}_1$ and $\bar{\nu}_2$ in units of 
$10^{-5}$eV$^2$, $L$~is
the distance traveled through the Earth in units of 1000~km, and $E$
is the neutrino energy in MeV.  We have assumed a constant matter
density inside the Earth, which is a good approximation for 
$L<10000$ km, i.e.\ as long as the neutrinos do not pass through the 
core of the Earth.

\begin{figure}[ht]
\begin{indented}
\item[]
\vspace{0.1cm}
\epsfig{file=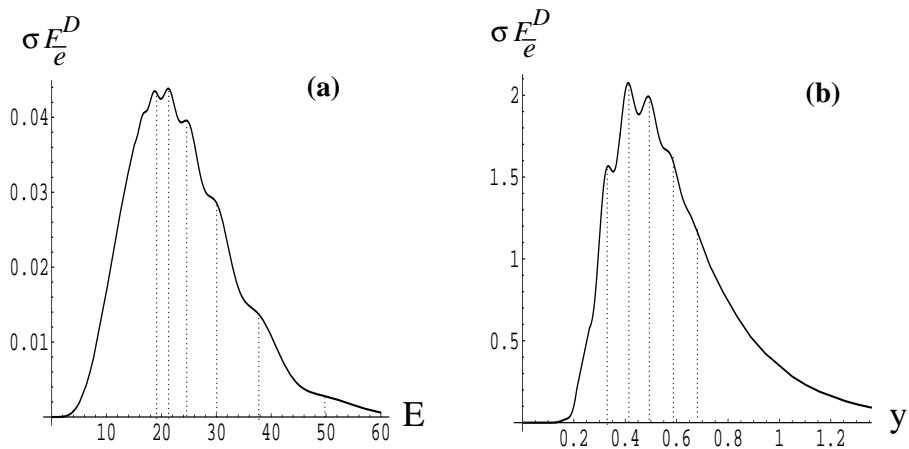,width=12cm}
\end{indented}
\caption{\label{fig:Ey} The energy spectrum (a) and the 
inverse-energy spectrum (b) of $\sigma F_\ebar^D$. 
The fluxes are normalized such that the area under each curve is unity.
For all the examples in this paper, we use the primary neutrino
flux parameters
$\alpha_\nuebar = \alpha_\nuxbar = 3.0$,
$\langle E_\nuebar \rangle = 15$ MeV, 
$\langle E_\nuxbar \rangle = 18$ MeV, 
$\Phi_\nuebar^0/\Phi_\nuxbar^0=0.8$,
which are realistic for the fluxes during the cooling phase.
For the mixing parameters, we use $\msqs = 6$ (in $10^{-5}$~eV$^2$) and
$\sin^2(2 \theta_\odot) =0.9$. The distance travelled through
the Earth is $L = 6$ (in 1000 km) unless otherwise specified.
}
\end{figure}

The energy dependence of $\bar{p}^D$ introduces modulations in the
energy spectrum of $\nuebar$. 
These modulations may be observed in the form of local peaks
and valleys in Fig.~\ref{fig:Ey}(a), which
shows the spectrum of the event rate, $\sigma F_\ebar^D$,
at a detector as a function of the neutrino energy.  
Fig.~\ref{fig:Ey}(b) shows the same neutrino signal as a
function of the ``inverse-energy'' parameter, defined as
\beq
y \equiv 12.5/E ~~.
\label{y-def}
\eeq
Whereas the distance between the peaks of the modulation increases
with energy in the energy spectrum, the peaks in the 
inverse-energy spectrum are nearly equispaced and hence have
a single dominating frequency. 
This makes it easier to distinguish these modulations 
from random background fluctuations that have no fixed pattern.

The equidistant peaks in the modulation of the inverse-energy spectrum
are a necessary feature of the Earth effects. Indeed,
the net $\nuebar$ flux at the detector may be written using (\ref{feDbar}) 
and (\ref{pbar}) in the form
\beq
F_{\bar{e}}^D =  \sin^2 \theta_{12}  F_\xbar^0 + 
\cos^2 \theta_{12} F_{\bar{e}}^0 + \Delta F^0
\bar{A}_\oplus \sin^2(\overline{\dmsq_\oplus} L y)~~,
\label{feDbar-y}
\eeq
where $\Delta F^0 \equiv (F_{\bar{e}}^0 - F_\xbar^0)$ depends
only on the primary neutrino spectra, whereas
$\bar{A}_\oplus \equiv - \sin 2\bar{\theta}_{e2}^\oplus
~ \sin (2\bar{\theta}_{e2}^\oplus - 2\theta_{12})$ 
depends only on the mixing parameters and is independent of the
primary spectra. 
The last term in (\ref{feDbar-y}) is the
Earth oscillation term that contains a frequency $k_\oplus \equiv
2 \dmbar L$ in $y$, with the coefficient $\Delta F^0 \bar{A}_\oplus$
being a comparatively slowly varying function of $y$.
The first two terms in (\ref{feDbar-y}) are also slowly varying 
functions of $y$, and hence contain frequencies in $y$ that are much 
smaller than $k_\oplus$.
The dominating frequency $k_\oplus$ is the one that appears in the 
modulation of the inverse-energy spectrum in Fig.~\ref{fig:Ey}(b).

The frequency $k_\oplus$ is completely independent of the primary
neutrino spectra, and indeed can be determined to a good accuracy from
the knowledge of the solar oscillation parameters, the Earth matter 
density, and the direction of the SN.
If this frequency component is isolated from the inverse-energy
spectrum of $\nuebar$, the Earth effects would be identified. 
In the next section, we shall show how this can be achieved through 
the Fourier transform of the inverse-energy spectrum.

\section{Identifying the Earth matter effects}
\label{identify}

\subsection{Fourier transform of the inverse-energy spectrum}
\label{ft}

Taking the Fourier transform of a function is the standard way of
extracting components of different frequencies present in the
function. The Fourier transform of a function $f(y)$ is
\beq
g(k) = \int_{-\infty}^{\infty} f(y) e^{i k y} dy
\label{ft-def}
\eeq
while the ``power spectrum'' $G_f(k) \equiv |g(k)|^2$ 
gives the strength of the frequency $k$ present in $f(y)$.

\begin{figure}[ht]
\begin{indented}
\item[]
\epsfig{file=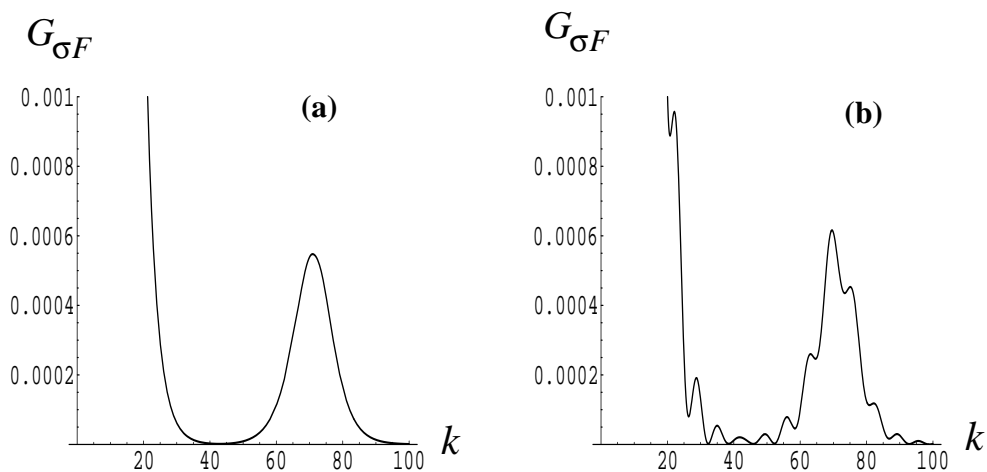,width=13cm}
\end{indented}
\caption{\label{fig:cutoff} The power spectrum $G_{\sigma F}(k)$
of $\sigma F_\ebar^D$ (a) and the same power spectrum with
a sharp energy threshold of 10 MeV (b).
}
\end{figure}

In Fig.~\ref{fig:cutoff}(a), we show the power spectrum $G_{\sigma F}(k)$
of the $y$-spectrum $\sigma F_\ebar^D$. The peak at 
$k \approx 2 \msqs L = 72$ 
corresponds to the oscillations in Earth matter. The large peak at low
values of $k$, which has the value of unity at $k=0$, is the dominant 
contribution due to the first two terms
in (\ref{feDbar-y}). As may be observed, the Earth effect peak is cleanly
isolated from the dominant contribution.
The figure also allows us to put a lower bound on the value of
$k_\oplus$ for which the Earth effects will be detectable:
if $\msqs L <20$, the Earth effect peak will be lost in the
dominant low frequency peak. Physically speaking, this implies that
the neutrinos have to travel a minimum distance through the
Earth for the Earth effects to be detectable.
So this is not a limitation of this particular method, but a
general limitation on the identification of Earth effects.

The Earth effect peak has a finite width due to the finite
$y$-dependence of the coefficient of the oscillating term
$\Delta F^0 \bar{A}_\oplus$, and of $\dmbar$. 
The magnitude of the width is a measure of the 
$y$-dependence of these quantities,
whereas the area under the peak gives the 
total contribution of the oscillating Earth effect term.

\begin{figure}[ht]
\begin{indented}
\item[]
\vspace{0.5cm}
\epsfig{file=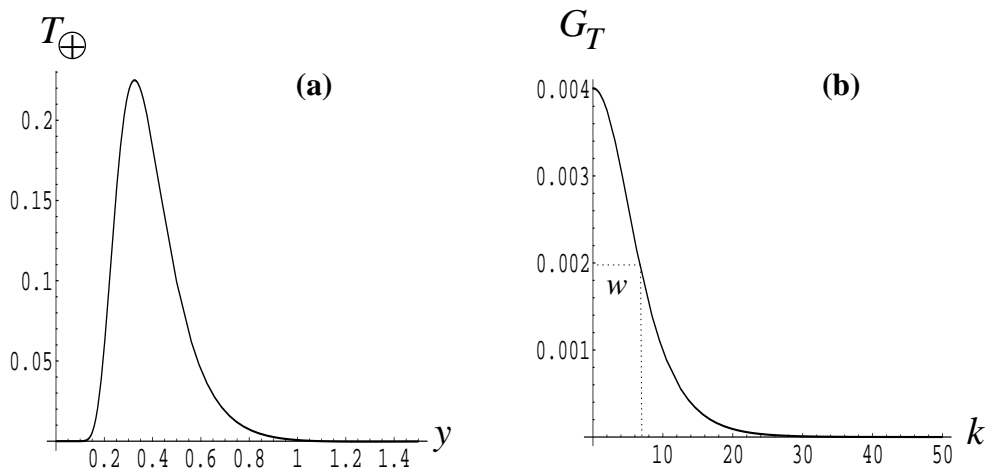,width=13cm}
\end{indented}
\caption{\label{fig:earthterm}
The Earth term $T_\oplus \equiv \sigma \Delta F^0 \bar{A}_\oplus$
as a function of $y$, and its power spectrum $G_T(k)$.
}
\end{figure}

In Fig.~\ref{fig:earthterm}(a), we show the envelope of the
Earth effect term,  $T_\oplus \equiv \sigma \Delta F^0 \bar{A}_\oplus$, 
for the typical neutrino spectra parameters during the cooling phase,
normalized such that $\int \sigma F_\xbar^0 dy =1$. 
The power spectrum of the Fourier transform $G_T(k)$ of this envelope
is shown in Fig.~\ref{fig:earthterm}(b). Since $T_\oplus$
is the coefficient of the Earth matter oscillations term 
$\sin^2(k_\oplus y)$ and $k_\oplus$ is nearly constant over all the
energy range, the width of the Earth effect peak in 
Fig.~\ref{fig:cutoff}(a) is due mainly to the width of
$G_T(k)$. We parameterize this width by $w$, defined as the value of
$k$ where $G_T(k)$ is half its maximum value. This would
correspond to the half width at half maximum of the $k_\oplus
\approx 72$ peak in Fig.~\ref{fig:cutoff}(a).

In  Fig.~\ref{fig:earthterm}(b), we observe $w \approx 7$. 
The value of $w$ depends on the primary fluxes, but it is in the
range of 3-10 for almost all the allowed parameter range. 
The exact value of $w$ is not crucial for this analysis, however
this estimation is useful in two ways. It allows us to get rid of
spurious peaks by applying a selection criterion of a minimum width
of the peak, and it also gives us the maximum range of $k$-values
around the mean $k_\oplus$ for which the Earth effect term
contributes.

The neutrino detectors have an energy threshold of
about 5--10 MeV, which corresponds to $y=$ 1.25--2.5. Although the
number of events below this energy threshold is expected to be very 
small (See Fig.~\ref{fig:Ey}), the presence of a sharp threshold may
introduce high frequency components in the $y$-spectrum. Although 
the threshold is not sharp in real life, we show an extreme example
of the effect of a sharp threshold in Fig.~\ref{fig:cutoff}(b),
taking a threshold of 10 MeV. 
Since the Fourier transform of a step function is a series of
equispaced frequencies, we observe that these high equispaced 
frequencies are superposed on top of the actual $G_{\sigma F}(k)$.
The spacing between these frequencies is much smaller than the
width of the Earth effect peak, so the observation of the peak 
and the area under the peak are not much affected, even in this 
extreme situation. In reality, the energy threshold is 
expected to be much smoother and should not affect the identification of
the peak and the measurement of its strength. In all the 
numerical simulations henceforth, we introduce a sharp energy threshold
of 5 MeV in order to take care of the threshold effects.

\subsection{The background due to statistical fluctuations}
\label{stat}

Since the spectrum is observed as a discrete set of neutrinos with 
individual energies (and hence, individual $y$ values), we have to
approximate $g(k)$ by the discrete sum
\beq
g(k) = \frac{1}{\sqrt{N}} \sum_{events} e^{i k y} ~~,
\label{ft-discrete}
\eeq
where $N$ is the total number of events. 
The statistical fluctuations in the spectrum can contribute to
frequency components that are comparable to $k_\oplus$. 
This acts as a background for the actual Earth effect peak.

The magnitude of this background can be estimated by assuming that
the distribution of $ky~({\rm modulo}~2\pi)$ is completely random.
This is definitely a good approximation for large values of $k$
independent of the details of the distribution of $y$.
In this approximation, the sum in
(\ref{ft-discrete}) above represents a two dimensional random
walk with unit step. This implies that the value of
$G_N(k)$ obeys an exponential distribution $P(G) = e^{-G}$,
which has a mean $\mu =1.0$ and variance $\sigma^2 =1.0$
independent of the number of events $N$.
The background due to the statistical fluctuations is 
thus nearly 1.0 on average, and the ``signal'' of the peak should 
rise above this average level in order to be identified.

\begin{figure}[ht]
\begin{indented}
\item[]
\epsfig{file=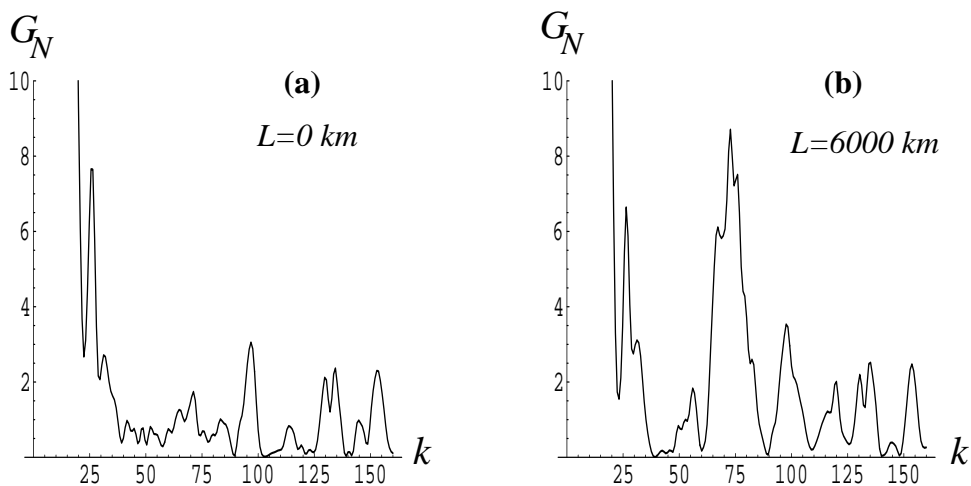,width=13cm}
\end{indented}
\caption{\label{fig:discrete} Background due to statistical fluctuations
when the Earth effects are absent (a) and the signal peak with the
Earth effects (b).
}
\end{figure}

This is illustrated in Fig~\ref{fig:discrete}(a), where the
distance travelled by the neutrinos through the Earth is
$L=0$. Note that the low frequency background is dominant
for $k<40$. 
In Fig.~\ref{fig:discrete}(b), we show the same neutrino
signal, but now with $L=6$. The Earth effect peak 
stands out above the background.
Though the level of the background is not affected by the
number of events, the magnitude of the signal grows 
proportional to $N$. 
Clearly, the Earth effect peak would grow above the 
background level when the number of events is sufficiently high.

Although the peak identification may be possible ``by eye'' in many cases, 
as in Fig~\ref{fig:discrete}(a), the standard
particle physics method of resonance identification may be employed
for the task.
Firstly, we apply a lower $k$-cut to cut away all $k<k_{\rm cut}$, 
where $k_{\rm cut}$
is the value of $k$ where the dominant peak falls to the background level.
In Fig~\ref{fig:discrete}(a), for example, $k_{\rm cut} \approx 40$.
Although we have made a rough estimate of the background above, the
actual background may be calculated by averaging the spectrum
$G_N(k)$ away from the value $k_\oplus$, i.e. in the region where
we do not expect the Earth effect peak.

We know from Fig.~\ref{fig:cutoff} that the half width at half
maximum of the peak in $k$ should be $w \approx 3$--$10$. We then expect 
that most of the Earth effects contribute to values of $k$ within a 
range of $W$ around the peak, where $W$ may be taken to be $4w$, 
for example.  We can then choose any interval in $k$ with 
a range of $W$, and integrate the power spectrum within this range. 
Applying the Central Limit Theorem to the background distribution
$P(G)$, we can estimate the integral of the background within this 
interval to be  $B = \mu_B W \pm \sigma_B \sqrt{W}$.
For $\mu_B=1.0$, $\sigma_B=1.0$ and $W=20$, we get $B \approx 20 \pm 4.5$.
This implies that $\int_W G_N(k)~ dk > 35$, for example, would
correspond to  a more than $3\sigma$ detection of a positive signal.
Depending on the actual data, the value of $W$ may be optimized.
The procedure of measuring the area under the peak is more efficient than
just measuring the peak height since it can weed out the spurious
high peaks that do not have the minimum width dictated by $G_T(k)$.

\begin{figure}[ht]
\begin{indented}
\item[]
\epsfig{file=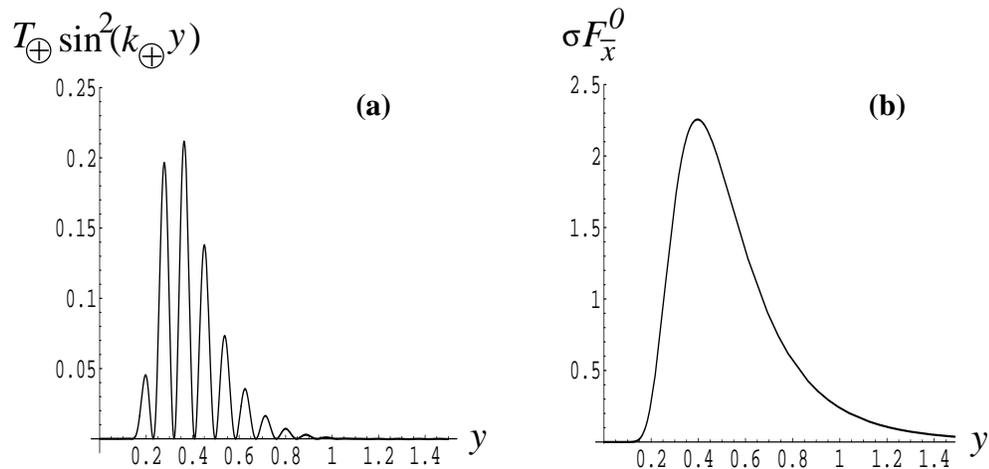,width=13cm}
\end{indented}
\caption{\label{fig:magnitude} The oscillating Earth effect term,
with the cross section of the inverse beta decay factored in (a).
It is normalized such that $\int \sigma F_\xbar^0=1$. The normalized
term $\sigma F_\xbar^0$ is shown in (b) for comparison. 
Note that the scale in (b) is 10 times the scale in (a). }
\end{figure}

The number of events needed in order to see the signal above the 
background depends strongly on the primary neutrino spectra,
since the coefficient of the Earth effect term depends on
$\Delta F^0$. 
In Fig.~\ref{fig:magnitude}(a), we show the Earth effect term 
$T_\oplus \sin^2(k_\oplus y)$ in comparison with the term
$\sigma F_\xbar^0$, normalized such that $\int \sigma F_\xbar^0=1$.
In Fig.~\ref{fig:magnitude}(b) we show the normalized 
term $\sigma F_\xbar^0$ for comparison. The area under this curve is unity.
The area ${\cal A}$ under the curve in Fig.~\ref{fig:magnitude}(a)
may then be taken to be a measure of the net contribution per event of the 
Earth effect term to the $y$-spectrum.
For the primary spectrum parameters used here, ${\cal A} \approx  0.03$ 
per event.

\begin{figure}[ht]
\begin{indented}
\item[]
\vspace{0.5cm}
\epsfig{file=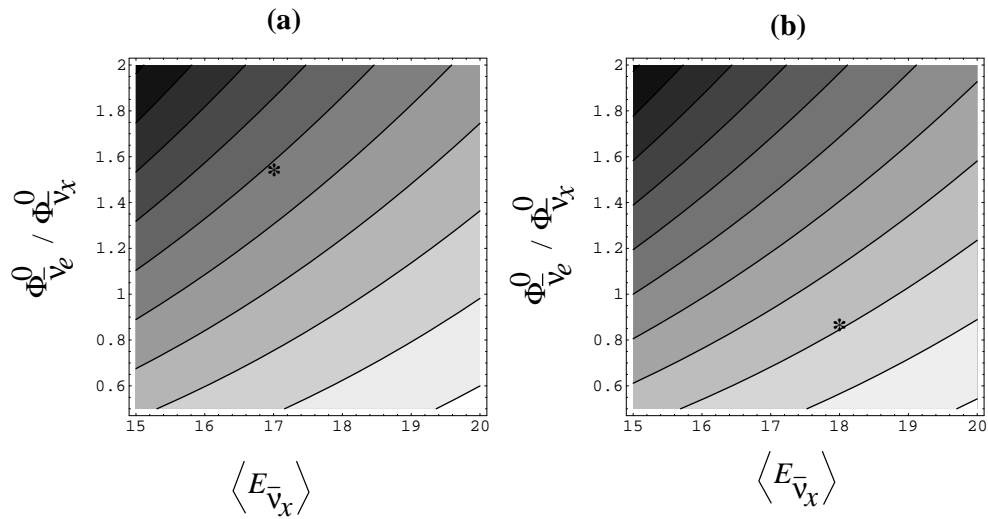,width=13cm}
\end{indented}
\caption{\label{fig:cont} The dependence of ${\cal A}$, the area under
the curve in Fig.~\ref{fig:magnitude}(a), on the primary neutrino flux 
parameters. The contour lines have values starting from 0.05 at the 
lightest end, with decrements of 0.01 going towards the darker
regions. The panel (a) uses the parameters $\langle E_\nuebar \rangle
=15$ MeV, $\alpha_\nuebar = 4$, $\alpha_\nuxbar=3$, panel (b) uses
$\langle E_\nuebar \rangle =15$ MeV, $\alpha_\nuebar = \alpha_\nuxbar=3$.
}
\end{figure}

We show in Fig.~\ref{fig:cont} the dependence of ${\cal A}$ 
on the parameters of the primary spectra. We keep the parameters
$\alpha_\nuxbar =3$ and $\langle E_\nuebar \rangle=15$
MeV fixed, and show ${\cal A}$ as a function of 
$\Phi_\nuebar^0/\Phi_\nuxbar^0$ and $\langle E_\nuxbar \rangle$
for two typical values of $\alpha_{\nuebar}$. 
During the accretion phase, $\alpha_\nuebar > \alpha_\nuxbar$,
which is the situation depicted in Fig.~\ref{fig:cont}(a).
During the cooling phase, the $\alpha$ values for both spectra
are very similar, which is depicted in Fig.~\ref{fig:cont}(b).
The stars roughly correspond to the parameter values obtained in the
SN simulations \cite{Noon03}. It may be observed that for
these parameter values, ${\cal A} \ll 0.01$ during the accretion 
phase and ${\cal A} \approx 0.03$ during the cooling phase, so that
the Earth effects during the accretion phase are expected to be much 
smaller. The relative magnitude of these effects thus also provides 
a test for the SN simulations.

We can estimate the number of events required for a peak identification
as follows. In order to get a $3\sigma$ identification, we need the
``signal'' contribution $S$ such that 
$S > 3 \sqrt{\sigma_S^2 + \sigma_B^2}$
where $\sigma_S^2 = S$. With $\sigma_B = 4.5$, this corresponds
approximately to $S>20$.  
In order to reach this strength $S$ with an average contribution of
${\cal A}\approx 0.03$ per event for example, we need at least 
$S/{\cal A} \approx 700$ events.
This is a very rough, and indeed an optimistic,
estimate of the minimum number of events needed
in order to detect Earth effects unambiguously.
In reality, the energy resolution of the detectors tend to
smear out the oscillations and decrease the magnitude of their strength,
thus increasing the required number of events.  
We shall study the effects of the energy resolution of the detectors
in Sec.~\ref{resol} and get a realistic estimate of the
number of events needed to identify the Earth effects.
The relative strength of the Earth effects for different primary 
spectra can still be read off from Fig.~\ref{fig:cont}.

Note that the above procedure can be employed without any prior 
knowledge of $k_\oplus$. Actually, if the value of $\dmbar$ and $L$ is
known, we already know the value of $k_\oplus$ to look for.
This helps in getting rid of any background due to spurious peaks.
On the other hand, if this peak is identified unambiguously, 
the value of $k_\oplus$ can help in improving the accuracy of
the measurement of $\msqs$. We shall study this in the next subsection.

\subsection{Determination of $\msqs$}
\label{dmsq-determine}

The current $3\sigma$ range of the solar mass squared difference 
$\msqs = (5.4$--$19) \times 10^{-5}$ eV$^2$ is obtained mainly 
through the combination of the limits from Super-Kamiokande 
and KamLAND. Although this range is expected to
narrow significantly with the future KamLAND data, it is worthwhile to
note that the Fourier analysis of the SN neutrino spectra can also
determine this value to an accuracy of a few percent.

Once the Earth effect peak is identified, the value of
$k_\oplus$ gives the value of $\dmbar$ since the value of $L$
should be well known once the SN direction is established.
The error in the measurement of the position of the peak may be
roughly estimated by $w/k_\oplus$. Since we expect $w\approx 3$--$10$,
and $k_\oplus$ may be as high as $2 \msqs \mbox{(max) } L \mbox{(max) }
\approx 400$, even this conservatively estimated error may be only
a few percent. 
As long as $k_\oplus > 40$, which is the minimum value of $k_\oplus$
for the Earth effects to be detectable, the error due to determination
of the peak position is less than 25\%.

\begin{figure}[ht]
\begin{indented}
\item[]
\vspace{0.5cm}
\epsfig{file=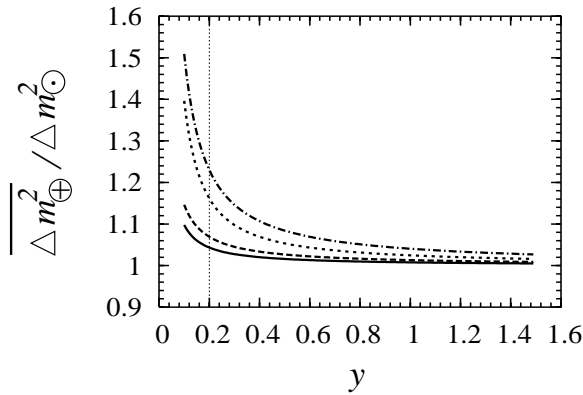,width=8cm}
\end{indented}
\caption{\label{fig:dmsq} The ratio $\dmbar/\msqs$ as a function of $y$
for different solar parameters. \\
Dot-dashed line: 
$\msqs = 6 \times 10^{-5}$ eV$^2$, $\sin^2(2\theta_{12}) = 0.7$,\\
Dotted line: 
$\msqs = 6 \times 10^{-5}$ eV$^2$, $\sin^2(2\theta_{12}) = 0.9$,\\
Dashed line: 
$\msqs = 18 \times 10^{-5}$ eV$^2$, $\sin^2(2\theta_{12}) = 0.7$,\\
Solid line: $\msqs = 18 \times 10^{-5}$ eV$^2$, $\sin^2(2\theta_{12}) = 0.9$.
}
\end{figure}

The value of $\dmbar$ is related to $\msqs$ by 
\beq
\dmbar = \msqs \left[ \sin^2 2\theta_\odot + 
(\cos 2\theta_\odot + 2~ V E/\msqs)^2 \right]^{1/2}
\label{dmbar-def}
\eeq
where $V$ is the magnitude of the matter potential inside the Earth. 
In Fig.~\ref{fig:dmsq}, we show the $y$-dependence of $\dmbar/\msqs$ with
various values of solar parameters. 
Since the $y$-spectrum is significant only for $y>0.2$,
the deviation of this ratio from unity at $y \approx 0.2$ may be taken
as a conservative estimate of the error on $\msqs$ from the SN
spectral analysis.
The figure shows that for $y>0.2$, the values of $\dmbar$ and $\msqs$ 
differ by less than 25\%, and this difference decreases with increasing 
$\msqs$ values.
Therefore, it may be safely assumed that effectively, we have 
$\dmbar \approx \msqs$ to within 25\%.

Throughout our analysis, we have assumed a constant matter density
inside the mantle of the Earth. Actually the density may vary by as
much as 30\% along the neutrino trajectory in the extreme case 
where the trajectory is nearly tangent to the core. 
Since $V \propto \rho$, this density variation contributes to the 
energy dependence of $\dmbar$ through the term involving $EV$ in 
(\ref{dmbar-def}), and smears the Earth effect peak.
However, since the variation in $E$ by a factor of five is already
taken care of in the above error estimation, the additional smearing
due to the Earth density variation may be safely neglected.

Combining the error due to the peak determination and the error due
to the difference in $\dmbar$ and $\msqs$, even conservatively 
the value of $\msqs$ would be known to within 35\%.
KamLAND may have already pinned this value down to a much better 
precision, however since no other planned experiment would
be able to reach a 35\% precision on $\msqs$ in near future, 
this could be the first confirmation of the $\msqs$ value 
measured at KamLAND. 
At high values of $\msqs$ and $L$, the error due to 
the peak width as well as the error due to the $\dmbar-\msqs$
difference decreases and the accuracy in the $\msqs$ measurement
may become competitive with KamLAND. For example, with
$\msqs = 15 \times 10^{-5}$ eV$^2$ and $L=3000$ km, 
the error in  $\msqs$ from the SN spectral analysis would be less
than 15\%, comparable with the accuracy expected at KamLAND
\cite{deGouvea:2001su}.

\section{The effect of finite energy resolution}
\label{resol}

The neutrino signal measured at the detector through the inverse
beta decay process $\nuebar p \to  n e^+$, 
which is the dominant reaction in both the water Cherenkov
as well as the scintillation detector, can be written in the form
\beq
N_e(E_e) \propto  \int  dE'_e ~{\cal R}(E_e, E'_e) {\cal E}(E'_e) 
\int dE~d\!\cos\theta\frac{d^2\sigma(E'_e,E, \cos\theta)}
{dE'_e~ d\!\cos\theta}  
F_\ebar(E)~,
\label{signal}
\eeq
where $E$ is the energy of the incoming antineutrino, $E'_e$ is the
true energy and $E_e$ is the measured energy of the outgoing positron.
The differential cross section of the inverse beta decay process is
$[d^2\sigma(E'_e,E,\cos\theta)/(dE'_e ~d\cos\theta)]$, where  
$\theta$ is the angle between the antineutrino and the positron.
We denote by ${\cal E}(E'_e)$ the efficiency of positron detection, and
${\cal R}(E_e, E'_e)$ is the energy resolution function.
We neglect the elastic
$\nu$--$e$ scattering interactions, which account for less than 
a few percent of the total number of events, and can even be 
weeded out through their strongly forward peaked angular 
distribution.

We take ${\cal E}(E'_e)$ to be uniform over the whole energy range except
for a sharp lower threshold of 5 MeV.
The resolution function 
${\cal R}(E_e, E'_e)$ is a Gaussian distribution with 
mean $E'_e$ and the standard deviation $\sigma_{\rm D}$ given
by the energy resolution of the detector D, where D is
SC for a scintillation detector and SK for a Cherenkov one.
We take the inverse beta decay process to be completely elastic,
so that the differential cross section 
involves a $\delta$ function nonvanishing only at
$E'_e \approx E/ [1 + (1 - \cos\theta) E/m_p]$, where $m_p$ is the
proton mass. 
The angular distribution is approximated to be isotropic, and the
cross section proportional to $E^2$.
This corresponds to the differential cross section in \cite{cross}
in the limit of degenerate neutron and proton masses, vanishing
electron mass, and isotropic angular distribution.
These approximations made here retain the essential features of the
observed spectra.
We finally normalize $N_e(E_e)$ to the number of
events using $\int N_e(E_e)~dE_e =N$.

The outgoing positrons are detected in the water Cherenkov detectors 
through the Cherenkov photons that they radiate. In the scintillation
detectors, the positrons are detected through photons produced in
the scintillation material. Since a larger number of photons can be 
produced in a scintillation detector, these have typically a much
better energy resolution than the water Cherenkov detectors. 
Indeed, whereas at Super-Kamiokande we have
$\sigma_{\rm SK}\approx 
1.5$~MeV~$(E/10$~MeV$)^{1/2}$ \cite{sk-res},
the energy resolution of a scintillation detector may be
as good as
$\sigma_{\rm SC}\approx~0.22$~MeV~$(E/10$~MeV$)^{1/2}$
\cite{lena},
which is more than a factor of 5 better.

\begin{figure}[ht]
\begin{indented}
\item[]
\epsfig{file=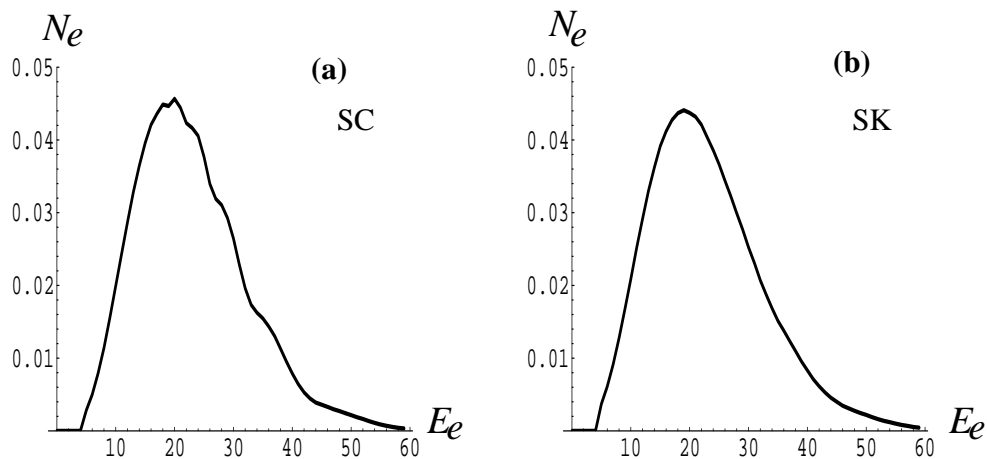,width=13cm}
\end{indented}
\caption{\label{compare-sksc} The energy spectrum as
observed at a scintillation detector (SC) and a 
Cherenkov detector (SK). 
All parameters except the energy resolution are identical.}
\end{figure}

Since the identification of Earth effects as described in this
paper relies on the detection of the modulations 
(see Fig.~\ref{fig:Ey}), and the finite energy resolution
tends to smear them out, it is clear that the 
energy resolution plays a crucial role in the 
efficiency of detecting Earth effects.
To illustrate this, we show in Fig.~\ref{compare-sksc}
the same energy spectrum of the signals as observed at 
a scintillation and a Cherenkov detector.
One can clearly see how the poor energy resolution of
the Cherenkov detector smears out the modulations.

\begin{figure}[ht]
\begin{indented}
\item[]
\epsfig{file=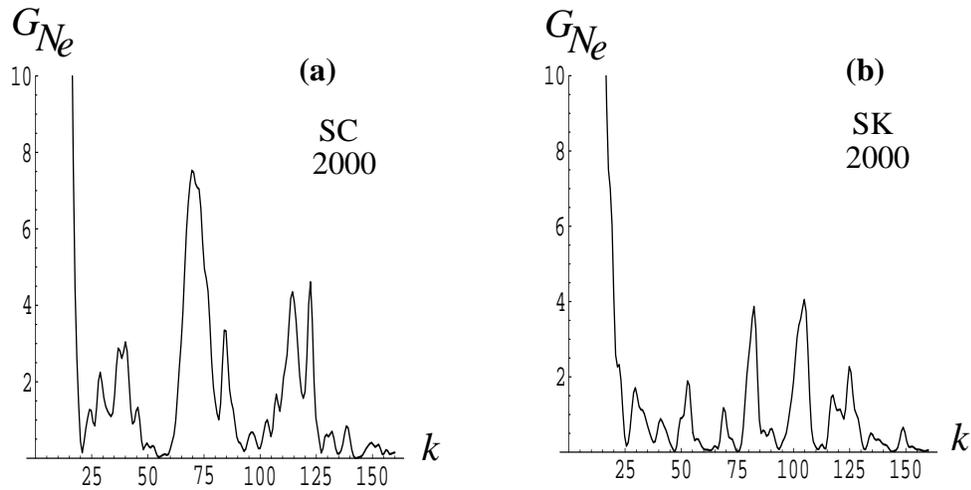,width=13cm}
\end{indented}
\caption{\label{fourier-2000} The Fourier transforms
of the simulated $y$-spectra in scintillation and 
Cherenkov detectors with 2000 events  each.}
\end{figure}

The difference between Figs.~\ref{compare-sksc}(a) and (b) is
reflected in the number of events required for the
signal to rise above the background at the two detectors.
A numerical simulation that generates inverse beta decay
events in each of the detectors illustrates the
comparative efficiency of the Fourier analyses at these two detectors.
In Fig.~\ref{fourier-2000}, we show the Fourier
transforms of the spectra at these two detectors with 2000
events each. Whereas the peak can be identified even by
eye at the scintillation detector, the Fourier power spectrum at
the Cherenkov detector is indistinguishable from background.

In  Fig~\ref{fourier-2000}(a), the area under the peak is around  80.
Since the area under the peak grows nearly linearly with the number
of events, this indicates that at a scintillation 
detector, the signal starts becoming visible at around $N \approx 1000$.
At the proposed 50 kiloton scintillation detector LENA \cite{lena} for
example, one expects about 13,000 events from a SN at 10 kpc.
This is an order of magnitude more than the statistics required for
identifying the Earth effects. Indeed, a much smaller scintillation
detector may be sufficient for this purpose.

\begin{figure}[ht]
\begin{indented}
\item[]
\epsfig{file=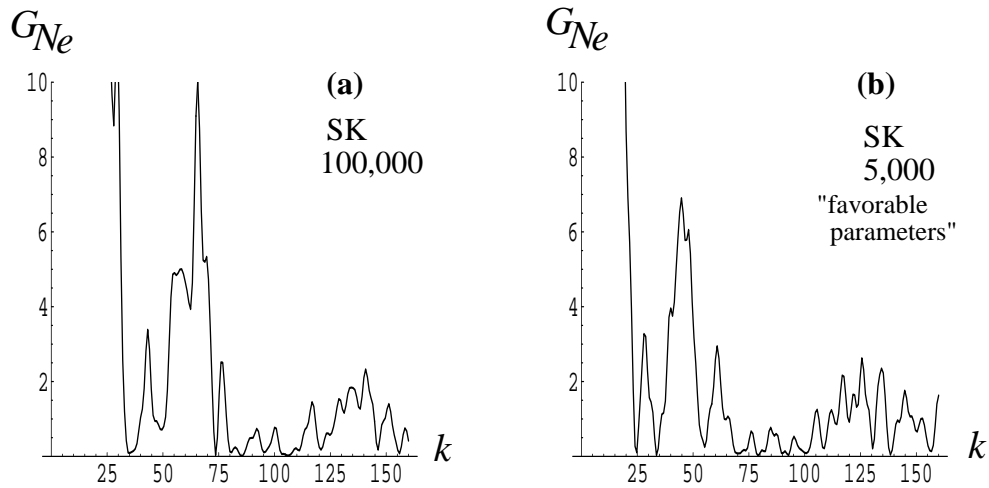,width=13cm}
\end{indented}
\caption{\label{sk-100K} The Fourier transforms
of the simulated $y$-spectra in a Cherenkov detector.
(a) uses 100,000 events and the typical neutrino spectral parameters, 
while (b) uses more extreme ``favorable'' parameters,
$\langle E_\nuxbar \rangle=21$ MeV and $L=4000$ km.
In the latter case, only 5,000 events are enough for the 
identification of Earth effects.
}
\end{figure}

At a water Cherenkov detector, the energy resolution is poor 
compared to a scintillation one. Indeed, the energy resolution is 
of nearly the same size as the wavelength of the Earth effect
modulations for $k_\oplus \sim 50$. It is therefore much harder to
identify the modulations. 
In Fig.~\ref{sk-100K}(a), we show the 
Fourier transform of the $y$-spectra generated at a Cherenkov detector
with 100,000 events. We expect that we need
more than 60,000 events to be able to identify the peak 
unambiguously. This indicates that Super-Kamiokande may be
too small for detecting the Earth effects by itself in this
parameter range.
However the proposed Hyper-Kamiokande would have the required
size.

If the modulation wavelength is larger than the energy resolution, 
which is the case for low $k_\oplus$, the peak is easier to detect. 
In addition, if the differences in the primary spectra of $\nuebar$ and 
$\nuxbar$ are larger than those taken in this paper till now, 
the peak identification can be achieved with a much smaller
number of events. 
In Fig.~\ref{sk-100K}(b), we show the power spectrum with the
average energies of the primary spectra differing by 40\%, and
the value of $L$ decreased to 4000 km. In such a ``favorable''
range, even a few thousand events would be enough at a water Cherenkov
detector, so that Super-Kamiokande has a chance of performing
a successful Earth effect detection. 
However in the light of the latest calculations of SN spectra,
such a large energy difference is basically ruled out.

Note that even for a scintillation detector, the energy resolution
becomes comparable to the modulation wavelength for large $k_\oplus$ 
values, so the number of events required will increase with
$k_\oplus$. At values of $k_\oplus > 200$, the modulations get
significantly washed out even at a scintillation detector.

\section{Conclusions}
\label{concl}

The modulations introduced in the neutrino spectrum by the 
Earth matter effects provide a way of detecting the presence of
these effects without prior assumptions about the 
flavor-dependent source spectra. 
We have demonstrated that a Fourier analysis of the inverse-energy 
spectrum of the $\nuebar$ signal may reveal a peak corresponding to the 
neutrino oscillation frequency in Earth.
The position of this peak is insensitive to the primary
spectra, and depends only on $\msqs$ and the
distance travelled through the Earth. 
We study the feasibility of the identification of this peak.

The number of events required for an unambiguous identification
depends crucially on the energy resolution of the detector.
The task is certainly feasible at a large water Cherenkov 
detector like Hyper-Kamiokande that can detect nearly $10^5$
events from a galactic SN. 
However, scintillation detectors generically have a much better
energy resolution than the Cherenkov ones,
and typically a scintillation detector 
can detect the Earth effects with only about 2000 events.
This indicates that a scintillation detector
large enough to detect a few thousand events from a galactic SN
is highly desirable.

The identification of the Earth effects severely restricts the 
neutrino mixing parameter space, since the effects are present 
only with certain combinations of the neutrino mass hierarchy and 
the mixing angle $\theta_{13}$. In particular, if 
$\sin^2 \theta_{13}$ is measured at a laboratory experiment
to be greater than $10^{-3}$, then the Earth effects on the $\nuebar$
spectrum imply the normal mass hierarchy. 
However if the Earth effects are not detected, it does not
rule out any neutrino mixing parameters, owing to
the current uncertainties in the primary fluxes.

A galactic SN is a rare event, expected to occur only a
few times per century. However this time scale should be compared with
that of those laboratory experiments which
would be sensitive to the mass hierarchy and values of 
$\sin^2 \theta_{13}$ as low as $10^{-3}$.
Determination of these two quantities is a difficult 
challenge even for the experiments
that may be running in the next few decades 
\cite{Huber:2002mx,Apollonio:2002en,Huber:2003pm}.
If a SN is observed within the next decade,
the information gained may even be useful
in fine-tuning the design parameters of these experiments.

The Fourier analysis described in this paper also yields the
value of $\msqs$ to a good accuracy if the matter effects are identified.
Though KamLAND should pin down $\msqs$ to within a few percent
in the next few years, a galactic SN will provide a completely
independent confirmatory test of this value.
For large values of $\msqs$ and $L$, the accuracy in $\msqs$
through the SN spectral analysis becomes comparable with
KamLAND.

Models of SNe generically predict time variations in the neutrino
flux spectra and hence in the magnitude of Earth effects.
The relative strength of the detected Earth effects as a function 
of time thus provides a test for these models.

In order to reap the benefits of the Earth effects, the detector
needs to be shadowed by the Earth from the SN. The probability of such an
occurrence is increased by having more than one detector
spaced far apart from each other. From the point of view of
SN neutrinos, several ``small'' (a few kiloton) scintillation detectors 
at different locations are far more useful than a single large one.

\section*{Acknowledgements}

This work was supported, in part, by the Deutsche
Forschungsgemeinschaft under grant No.\ SFB-375 and by the European
Science Foundation (ESF) under the Network Grant No.~86 Neutrino
Astrophysics. 
We thank M. Kachelriess for a careful reading of the 
manuscript, E. Lisi for helpful comments, and
R. Tom\`as for useful discussions.

\section*{References}


\end{document}